\documentclass[aps,prl,showpacs,twocolumn,superscriptaddress]{revtex4}
\usepackage{amsmath}
\usepackage[colorlinks]{hyperref}
\usepackage{amssymb}
\usepackage{epsfig}
\usepackage{color}
\usepackage{graphicx,amsmath}
\hypersetup{colorlinks,citecolor=red,linkcolor=blue,urlcolor=blue}

\begin{document}
\title{Revealing quantum spin liquid in
the herbertsmithite $\rm ZnCu_{3}(OH)_6Cl_{2}$}
\author{V. R. Shaginyan}\email{vrshag@thd.pnpi.spb.ru} \affiliation{Petersburg
Nuclear Physics Institute of NRC "Kurchatov Institute",
Gatchina, 188300, Russia}\affiliation{Clark Atlanta University,
Atlanta, GA 30314, USA} \author{M. Ya. Amusia}\affiliation{Racah
Institute of Physics, Hebrew University, Jerusalem 91904,
Israel}\affiliation{Ioffe Physical Technical Institute, RAS, St.
Petersburg 194021, Russia}
\author{J.~W.~Clark}
\affiliation{McDonnell Center for the Space Sciences \&
Department of Physics, Washington University, St.~Louis, MO
63130, USA} \affiliation{Centro de Ci\^encias Matem\'aticas,
Universidade de Madeira, 9000-390 Funchal, Madeira, Portugal}
\author{G. S. Japaridze}\affiliation{Clark Atlanta
University, Atlanta, GA 30314, USA}
\author{A. Z.
Msezane}\affiliation{Clark Atlanta University, Atlanta, GA
30314, USA}\author{V. A. Stephanovich}\affiliation{Institute of
Physics, Opole University, Oleska 48, 45-052, Opole, Poland}
\author{Y. S.
Leevik}\affiliation{National Research University Higher School
of Economics, St.Petersburg, 194100, Russia}
\author{E. V. Kirichenko}\affiliation{Institute of Mathematics and
Informatics,Opole University, Oleska 48, 45-052, Opole, Poland}

\begin{abstract}

Based on experimental data and our theoretical analysis, we
provide a strategy for unambiguous establishing of gapless
quantum spin liquid state (QSL) in herbertsmithite and other
materials. To clarify the nature of QSL, we recommend
measurements of heat transport, low-energy inelastic neutron
scattering and optical conductivity under the application of
external magnetic field at low temperatures. We also suggest
that artificially introduced inhomogeneity into $\rm
ZnCu_{3}(OH)_6Cl_2$ can stabilize QSL, and serves as a test
elucidating the contribution coming from impurities. We predict
the results of these measurements in the case of gapless QSL.
\end{abstract}

\pacs{64.70.Tg, 75.40.Gb, 78.20.-e, 71.10.Hf}

\maketitle

In a geometrically frustrated magnet, spins are prevented from
forming an ordered alignment, so that even at temperatures close
to absolute zero they collapse into a liquid-like state called a
quantum spin liquid (QSL). The herbertsmithite $\rm
ZnCu_3(OH)_6Cl_2$ has been exposed as a $S=1/2$ kagome
antiferromagnet, and recent experimental investigations have
revealed its unusual behavior
\cite{helt,herb2,herb3,herb,t_han:2012,t_han:2014}. The
electrostatic forces balance for $\rm Cu^{2+}$ ions in a kagome
structure makes them to occupy the distorted octahedral sites.
In the herbertsmithite structure, magnetic planes formed by $\rm
Cu^{2+}$ $S=1/2$ ions are interspersed with nonmagnetic $\rm
Zn^{2+}$ layers. In samples, $\rm Cu^{2+}$ defects occupy the
nonmagnetic $\rm Zn^{2+}$ sites between the kagome layers with
$x\simeq 15\%$ probability, thus introducing randomness and
inhomogeneity into the lattice \cite{Han}. We suggest that the
influence of this inhomogeneity on the properties of $\rm
ZnCu_3(OH)_6Cl_2$ facilitates the frustration, and can be tested
by varying $x$, as it is observed in measurements on the
verdazyl-based complex $\rm Zn(hfac)_2(A_xB_{1-x})$
\cite{screp-17}. The experiments made on $\rm ZnCu_3(OH)_6Cl_2$
did not find any traces of magnetic order in it. Neither they
have found the spin freezing down to temperatures of around 50
mK. This shows that the herbertsmithite is the best candidate to
contain the above QSL.
\cite{helt,herb2,herb3,herb,t_han:2012,t_han:2014}. These
results are confirmed by model calculations indicating that the
ground state of kagome antiferromagnet is a gapless spin liquid
\cite{prr,shaginyan:2011,shaginyan:2012:A,shaginyan:2011:C,shaginyan:2013:D,book,Normand}.
On the other hand, it is shown that the intrinsic local spin
susceptibility $\chi_{\rm kag}$ vanished above $T\geq 10$ K and
magnetic fields $B\geq 10$ T \cite{sc_han}. It has recently been
suggested that there can exist a small spin-gap in the kagome
layers \cite{Han,Han11,sc_han}, see also Refs \cite{norman,zhou}
for recent review. The results reported are based on both
experimental facts and their theoretical interpretation in the
framework of impurity model \cite{Han,Han11,norman}. The
experimental data are derived from high-resolution low-energy
inelastic neutron scattering on $\rm ZnCu_{3}(OH)_6Cl_2$
single-crystal. The impurity model assumes that the
corresponding ensemble may be represented as a simple cubic
lattice in the dilute limit below the percolation threshold. The
model then suggests that in the absence of magnetic fields the
bulk spin susceptibility $\chi$ exhibits a divergent Curie-like
tail, indicating that some of the $\rm Cu$ spins act like weakly
coupled impurities \cite{Han,Han11,norman}. The same behavior is
recently reported in a new kagome quantum spin liquid candidate
$\rm Cu_3Zn(OH)_6FBr$ \cite{feng}. As a result, we observe a
challenging contradiction between two sets of experimental data
when some of them state the absent of a gap, while the other
present evidences in the favor of gap.

Main goal of our letter is to attract attention to experimental
studies of $\rm ZnCu_3(OH)_6Cl_2$ that can unambiguously reveal
both the physics of QSL and the existence, or absence, of a
possible gap in spinon excitations that form the thermodynamic,
transport and relaxation properties. To unambiguously clarify
the nature of QSL in herbertsmithite, we recommend the
measurements of heat transport, low-energy inelastic neutron
scattering and optical conductivity $\overline{\sigma}$ in $\rm
ZnCu_{3}(OH)_6Cl_2$ subjected to external magnetic field. We
suggest that the influence of impurities on the properties of
$\rm ZnCu_3(OH)_6Cl_2$ can be tested by varying $x$. We predict
results of these measurements.

To analyze the QSL properties theoretically, we employ the
strongly correlated quantum spin liquid (SCQSL) model
\cite{shaginyan:2011,prr,shaginyan:2011:C}. A simple kagome
lattice may have a dispersionless topologically protected branch
of the quasiparticle spectrum with zero excitation energy, that
is the so-called flat band
\cite{shaginyan:2011,prr,green,jltp:2017}. In that case a
topological fermion condensation quantum phase transition
(FCQPT) can be considered as QCP of the $\rm ZnCu_3(OH)_6Cl_2$,
with SCQSL is composed of heavy fermions, or spinons, with zero
charge and effective mass $M^*$, occupying the corresponding
Fermi sphere with the Fermi momentum $p_F$. Consequently, the
properties of insulating magnets coincide with those of
heavy-fermion metals with one exception: Namely, typical
insulator resists the electric current
\cite{prr,shaginyan:2011,shaginyan:2012:A,shaginyan:2011:C,book,jltp:2017}.

At $B=0$, contrary to the Landau Fermi liquid (LFL) behavior,
where the effective mass $M^*$ is approximately constant, this
quantity becomes strongly temperature dependent, demonstrating
the non-Fermi liquid (NFL) behavior \cite{prr}
\begin{equation}
M^*(T)\simeq a_TT^{-2/3}.\label{MTT}
\end{equation}
At finite $T$, the system transits to the LFL behavior, being
subjected to the magnetic field
\begin{equation}
M^*(B)\simeq a_BB^{-2/3}.\label{MBB}
\end{equation}

\begin{figure} [! ht]
\begin{center}
\includegraphics [width=0.47\textwidth]{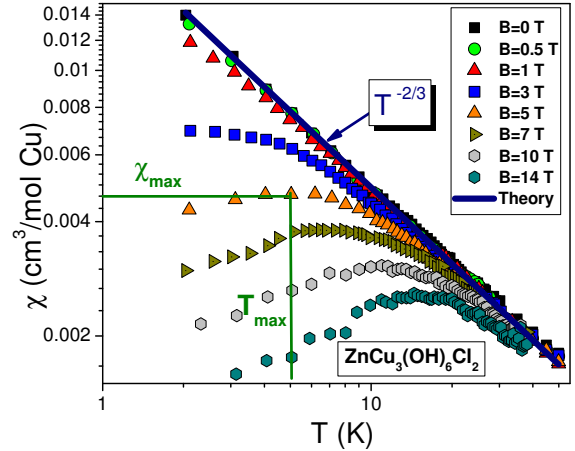}
\end{center}
\vspace*{-0.8cm} \caption{(Color online) Measured temperature
dependence of the magnetic susceptibility $\chi$ of $\rm
ZnCu_3(OH)_6Cl_2$ from Ref.~\cite{herb3} at magnetic fields
shown in the legend. Illustrative values of $\chi_{\rm
max}\propto M^*_{M}$ and $T_{\rm max}$ at $B=3$ T are also
shown. A theoretical calculation at $B=0$ is plotted as the
solid curve, which represents $\chi(T)\propto T^{-\alpha}$ with
$\alpha=2/3$ \cite{shaginyan:2011,shaginyan:2011:C,book}.}
\label{fig01}
\end{figure}

The introduction of "internal" (or natural) scales greatly
simplifies understanding the thermodynamic, transport and
relaxation properties \cite{prr}. Namely, near FCQPT the
effective mass $M^*(B,T)$ reaches its maximal value $M^*_M$ at
certain temperature $T_{M}\propto B$. Hence, to measure the
effective mass and temperature, it is convenient to introduce
the scales $M^*_M$ and $T_{M}$ respectively, see Fig.
\ref{fig01}. This generates the normalized effective mass
$M^*_N=M^*/M^*_M$ and the temperature $T_N=T/T_{M}$. Near FCQPT
the normalized effective mass $M^*_N(T_N)$ can be well
approximated by a simple universal interpolating function. The
interpolation occurs between the LFL and NFL states, reflecting
the universal scaling behavior of $M^*_N$ \cite{prr}
\begin{equation}M^*_N(y)\approx c_0\frac{1+c_1y^2}{1+c_2y^{8/3}}.
\label{UN2}
\end{equation}
Here, $y=T_N=T/T_{M}$, $c_0=(1+c_2)/(1+c_1)$, where $c_1$ and
$c_2$ are free parameters. The magnetic field $B$ enters only in
the combination $\mu_BB/T$, making $T_{M}\sim \mu_BB$. It
follows from Eq.~\eqref{UN2} that
\begin{equation}
\label{TMB} T_M\simeq a_1\mu_BB,
\end{equation}
where $a_1$ is a dimensionless factor, $\mu_B$ is the Bohr
magneton. Thus, in the presence of magnetic field the variable
$y$ becomes $y=T/T_{M}\sim T/\mu_BB$. Expression \eqref{TMB}
permits to conclude that Eq. \eqref{UN2} describes the scaling
behavior of the effective mass as a function of $T$ and $B$: The
curves $M^*_{N}$ at different magnetic fields $B$ merge into a
single one in terms of the normalized variable $y=T/T_M$. Since
the variables $T$ and $B$ enter symmetrically, Eq. \eqref{UN2}
also describes the scaling behavior of $M^*_{N}(B,T)$ as a
function of $B$ at fixed $T$.

\begin{figure} [! ht]
\begin{center}
\includegraphics [width=0.47\textwidth]{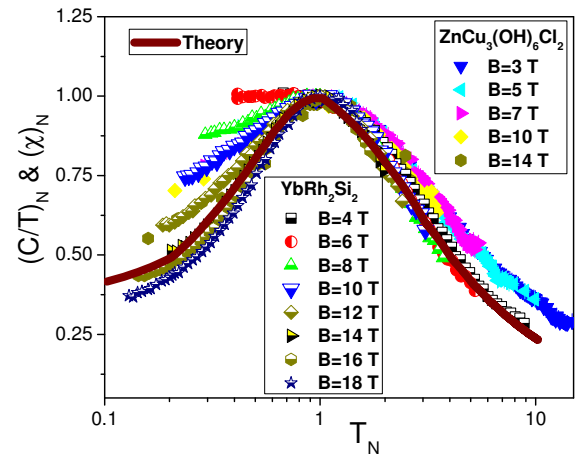}
\end{center}
\caption{Normalized susceptibility $\chi_N=\chi/\chi_{\rm
max}=M^*_N$ versus normalized temperature $T_N=T/T_{\rm max}$
(see Eq. \eqref{UN2} and Fig. \ref{fig01}) extracted from the
measurements of the magnetic susceptibility $\chi$ in magnetic
fields $B$ on $\rm ZnCu_3(OH)_6Cl_2$ \cite{herb3} shown in Fig.
\ref{fig01}. Normalized specific heat $(C/T)_N=M^*_N$ is
extracted from the measurements of the specific heat $C/T$ on
$\rm YbRh_2Si_2$ in magnetic fields $B$ \cite{steg1}. The
corresponding values of $B$ are listed in the legends. Our
calculations made at $B\simeq B^*$ when the quasiparticle band
is fully polarized are depicted by the solid curve tracing the
scaling behavior of $M^*_N$
\cite{book,shaginyan:2011:C}.}\label{fig4_1}
\end{figure}

To examine the impurity model, we first refer to the
experimental behavior of the magnetic susceptibility $\chi$ of
herbertsmithite.  It is seen from Fig.~\ref{fig01} that the
magnetic susceptibility diverges $\chi(T)\propto T^{-2/3}$ in
magnetic fields $B\leq 1$ T
\cite{herb3,shaginyan:2011,shaginyan:2011:C}, as shown by the
solid line. In the case of weakly interacting impurities it is
suggested that the low-temperature behavior of $\chi_{\rm
CW}(T)\propto 1/(T+\theta)$ can be approximated by a Curie-Weiss
law \cite{Han,Han11,sc_han}, with $\theta$ being a vanishingly
small Weiss temperature. However, given that $\chi(T)\propto
T^{-2/3}$, the Curie-Weiss approximation is in conflict with
both experiment \cite{herb3} and theory
\cite{shaginyan:2011,shaginyan:2011:C,book}. Moreover, it is
seen from Fig. \ref{fig4_1} that normalized magnetic
susceptibility $\chi$ behaves like the normalized heat capacity,
extracted from measurements on $\rm YbRh_2Si_2$ in high magnetic
fields \cite{steg1}, and does not exhibit any gap. This
observation confirms the absent of the spin gap in $\rm
ZnCu_3(OH)_6Cl_2$ and invalidity of separating the contributions
coming from the impurities. While $\chi_{\rm kag}$ obeys
$\chi_{\rm kag}(T)=\chi(T) -\chi_{\rm CW}(T)$, leading to
$\chi_{\rm kag}(T\to0)\to 0$ and to the erroneous claim that a
gap has been observed \cite{Han11,sc_han}. To explain the
observed behavior of $\chi$, defining the real thermodynamic
properties of $\rm ZnCu_3(OH)_6Cl_2$, it is necessary to
consider the impurities and the kagome planes as an integral
entity that forms QSL
\cite{prr,shaginyan:2011,shaginyan:2012:A,shaginyan:2011:C,
shaginyan:2013:D,book}.
\begin{figure} [! ht]
\begin{center}
\includegraphics [width=0.47\textwidth]{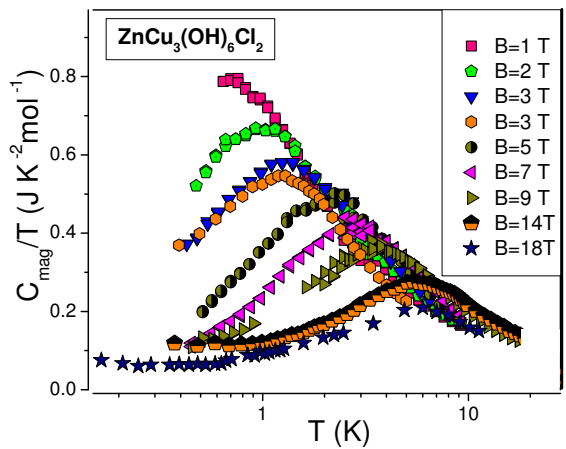}
\end{center}
\vspace*{-0.8cm} \caption{(Color online) Specific heat $C_{\rm
mag}/T$ measured on powder \cite{helt,herb2} and single-crystal
\cite{herb,t_han:2012,t_han:2014} samples of herbertsmithite is
displayed as a function of temperature $T$ for fields $B$ shown
in the legend.} \label{fig02}
\end{figure}
It is also suggested that measurements of low energy inelastic
neutron scattering on single crystals of herbertsmithite allow
one to subtract the impurity scattering $S_{\rm imp}(\omega)$
from the total scattering $S_{tot}(\omega)$ to obtain a measure
of the intrinsic scattering $S_{kag}(\omega)=S_{tot}(\omega)
-aS_{\rm imp}(\omega)$, taking $a$ as a fitting parameter. As a
result, it is found that $S_{\rm kag}(\omega)\to 0$ as $\omega$
decreases below the energy of 0.7 meV, leading to the gap
existence. As we have shown above analyzing the magnetic
susceptibility, such a subtraction leads to the erroneous
conclusion that a gap has been found.

Let us consider somewhat further the inadequacy of the impurity
model and its conclusion about spin gap existence when
confronted with experimental findings. We see from
Fig.~\ref{fig01} that LFL behavior of the magnetic
susceptibility $\chi$ is demonstrated at least for $B\geq 3$ T
and low temperatures $T$. At such temperatures and magnetic
fields the impurities should become fully polarized. Thus,
assuming the impurities are fully polarized and hence do not
contribute to $\chi$, one has simply $\chi_{\rm kag}(T)
=\chi(T)$. Analogous behavior of the heat capacity follows from
Fig.~\ref{fig02}. LFL behavior of $C_{\rm mag}/T$ emerges under
the application of the same fields. Consequently, we may
conclude that at least at $B\geq 3$ T and $0.2\leq T\leq 2$ K,
the contributions to both $\chi$ and $C_{\rm mag}/T$ from the
impurities are negligible; rather, one expects them to be
dominated by the kagome lattice, exhibiting a spin gap in the
kagome layers \cite{Han,Han11,sc_han}. Thus, one would expect
both $\chi(T)$ and $C_{\rm mag}(T)/T$ to approach zero at $T
\leq 2$ K and $B\geq 3$ T. It is clear from Figs.~\ref{fig01},
\ref{fig4_1}, and \ref{fig02}, that this is not the case. These
conclusions agree with recent experimental findings that the
low-temperature plateau in local susceptibility identifies the
spin-liquid ground state as gapless one \cite{zorko}, while
recent theoretical analysis confirms the absence of a gap
\cite{Normand,prl_2017}. Moreover, we suggest that the growing
$x$, elevating randomness and inhomogeneity of the lattice and
therefore facilitating the frustration of the lattice, can
stabilize QSL. This observation can be tested in experiments on
samples of herbertsmithite with different $x$ under the
application of magnetic field.
\begin{figure} [! ht]
\begin{center}
\includegraphics [width=0.47\textwidth]{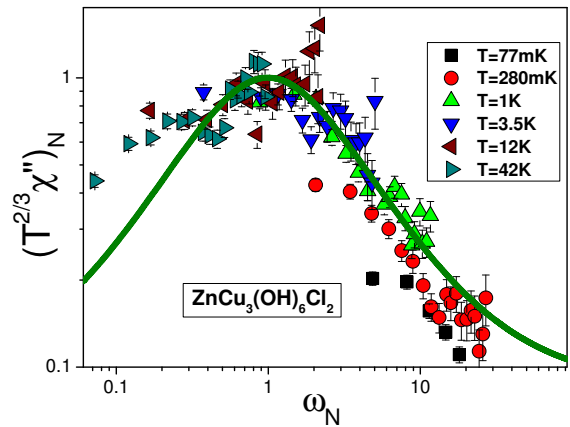}
\end{center}
\caption{(Color online) Scaling behavior of the normalized
dynamic spin susceptibility $(T^{2/3}\chi'')_N$. Data are
extracted from measurements on the herbertsmithite $\rm
ZnCu_3(OH)_6Cl_2$ \cite{herb3}. Solid curve: Theoretical
calculations based on Eq.~\eqref{SCHIN}
\cite{shaginyan:2012:A,book}.}\label{fig05}
\end{figure}

The same outcomes can be drawn from the results of
neutron-scattering measurements of the dynamic spin
susceptibility $\chi({\bf q},\omega,T) =\chi{'}({\bf
q},\omega,T)+i\chi{''}({\bf q},\omega,T)$ as a function of
momentum $q$, frequency $\omega$, and temperature $T$. Indeed,
these results play a crucial role in identifying the properties
of the quasiparticle excitations involved. At low temperatures,
such measurements reveal that the corresponding quasiparticles
-- of a new type insulator -- are represented by spinons, form a
continuum, and populate an approximately flat band crossing the
Fermi level \cite{Han:2012}. The imaginary part
$\chi''(T,\omega_1)$ satisfies the equation
\cite{shaginyan:2012:A,book}
\begin{equation}\label{SCHII}
T^{2/3}\chi''(T,\omega_1)\simeq\frac{a_1\omega_1}{1+a_2\omega_1^2},
\end{equation}
where $a_1$ and $a_2$ are constants and
$\omega_1=\omega/(T)^{2/3}$. It is seen from Eq. \eqref{SCHII}
that $T^{2/3}\chi''(T,\omega_1)$ has a maximum
$(T^{2/3}\chi''(T,\omega_1))_{\rm max}$ at some $\omega_{\rm
max}$ and depends on the only variable $\omega_1$. Equation
\eqref{SCHII} confirms the scaling behavior of $\chi'' T^{0.66}$
experimentally established in Ref. \cite{herb3}. Similar to Eq.
\eqref{UN2}, we introduce the dimensionless function
$(T^{2/3}\chi'')_{N}=T^{2/3}\chi''/(T^{2/3}\chi'')_{\rm max}$
and the (dimensionless) variable $\omega_N=\omega_1/\omega_{\rm
max}$. In this case, Eq. \eqref{SCHII} is modified to read
\begin{equation}\label{SCHIN}
(T^{2/3}\chi'')_N\simeq\frac{b_1\omega_N}{1+b_2\omega_N^2},
\end{equation}
with $b_1$ and $b_2$ are fitting parameters. Their role is to
adjust the function on the right hand side of Eq. \eqref{SCHIN}
to reach its maximum value 1 at $\omega_N=1$. In such a
situation it is expected that the dimensionless normalized
susceptibility $(T^{2/3}\chi'')_{N}=T^{2/3}\chi''
/(T^{2/3}\chi'')_{\rm max}$ exhibits scaling as a function of
the dimensionless energy variable $\omega_N $, as it is seen
from Fig. \ref{fig05}. We predict that if measurements of
$\chi''$ are taken at fixed $T$ as a function of $B$, then with
respect to Eq. \eqref{MBB}, we again obtain that the function
$B^{2/3}\chi''(\omega)$ exhibits the scaling behavior with
$\omega_N=\omega_1/\omega_{max}$
\begin{equation}\label{SCHB}
(B^{2/3}\chi'')_N\simeq\frac{d_1\omega_N}{1+d_2\omega_N^2},
\end{equation}
Similarly, $d_1$ and $d_2$ are fitting parameters adjusted such
that the function $(B^{2/3}\chi'')_{N}$ reaches unity at
$\omega_N=1$. If the system is exactly at a FCQPT point, the
above scaling is valid down to lowest temperatures. It would
also be crucial to carry out the measurements of low energy
inelastic neutron scattering on $\rm ZnCu_3(OH)_6Cl_2$ single
crystals under the application of relatively high magnetic
fields. Latter measurements permit to directly observe possible
gap since in this case the contribution from supposed impurities
is negligible, as we have seen above in the case of the spin
susceptibility $\chi$.

Measurements of  heat transport are particularly salient in that
they probe the low-lying elementary excitations of QSL in $\rm
ZnCu_3(OH)_6Cl_2$ and potentially reveal itinerant spinons that
are mainly responsible for the heat transport. Surely, the
overall heat transport is contaminated by the phonon
contribution; however, this contribution is hardly affected by
the magnetic field $B$. SCQSL in herbertsmithite behaves like
the electron liquid in HF metals -- provided the charge of an
electron is set to zero. As a result, the thermal resistivity
$w$ of the SCQSL is given by
\cite{shaginyan:2011:C,shaginyan:2013:D,book}
\begin{equation}\label{wr}
w-w_0=W_rT^2\propto\rho-\rho_0\propto(M^*)^2T^2,
\end{equation}
where $W_r{T^{2}}$ represents the contribution of spinon-spinon
scattering to thermal transport, being analogous to the
contribution $AT^2$ from electron-electron scattering to charge
transport, and $\rho$ is the longitudinal magnetoresistivity
(LMR). Also, $w_0$ and $\rho_0$ are the residual thermal and
electric resistivity respectively.
\begin{figure} [! ht]
\begin{center}
\includegraphics [width=0.52\textwidth]{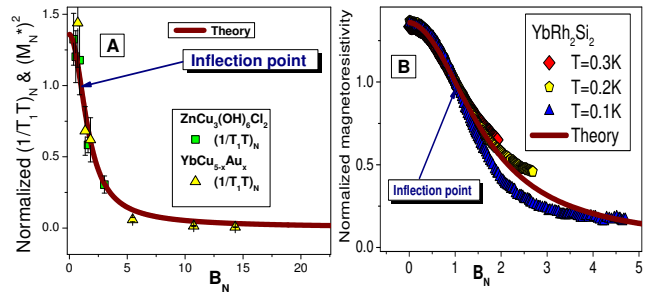}
\end{center}
\caption{(Color online) Panel (A). Normalized spin-lattice
relaxation rate $(1/T_1T)_N$ at fixed temperature as a
function of magnetic field. Data for $(1/T_1T)_N$ extracted
from measurements on $\rm ZnCu_3(OH)_6Cl_2$ are shown by
solid squares \cite{imai} and those extracted from
measurements on $\rm YbCu_{5-x}Au_{x}$ at $x=0.4$, by the
solid triangles \cite{carr}. The inflection point at which
the normalization is taken is indicated by the arrow. Panel
(B). Magnetic field dependence of the normalized
magnetoresistance $\rho_N$, extracted from LMR of $\rm
YbRh_2Si_2$ at different temperatures \cite{gegmr} listed
in the legend. The inflection point is shown by the arrow.
In both panels (A) and (B), the calculated result is
depicted by the same solid curve, tracing the scaling
behavior of $W_r\propto(M^*)^2$ (see Eqs.~\eqref{wr} and
\eqref{WT}).}\label{T12}
\end{figure}
Now, we consider the effect of a magnetic field $B$ on the
spin-lattice relaxation rate {$1/(T_1T)$}. Fig.~\ref{T12} A
shows the normalized spin-lattice relaxation rate $1/(T_1T)_N$
at fixed temperature versus magnetic field $B$. It is seen that
increasing of $B$ reduces progressively  $1/(T_1T)$. Also, the
curve in Fig.~\ref{T12} A  has an inflection point at some
$B=B_{\rm inf}$, marked by the arrow. To detect the scaling
behavior in this case, we normalize { $1/(T_1T)$} by its value
at the inflection point, while the magnetic field is normalized
by $B_{\rm inf}$.  Taking into account the relation
$1/(T_1T)_N\propto(M^*)^2$, we expect that a strongly correlated
Fermi system located near its QCP would exhibit the similar
behavior of { $1/(T_1T)_N$}
\cite{prr,shaginyan:2011:C,shaginyan:2013:D,book}.
Significantly, Fig.~\ref{T12} A shows that the herbertsmithite
$\rm ZnCu_3(OH)_6Cl_2$ \cite{imai} and the HF metal $\rm
YbCu_{5-x}Au_{x}$ \cite{carr} do exhibit the same behavior for
the normalized spin-lattice relaxation rate. As seen from
Fig.~\ref{T12} A for $B\leq B_{\rm inf}$ (or $B_N\leq1$) the
normalized relaxation rate $1/(T_1T)_N$ depends weakly on the
magnetic field, while it diminishes at the higher fields
\cite{prr,shaginyan:2011:C,shaginyan:2013:D,book} according to
\begin{equation}\label{WT}
W_r\propto{ 1/(T_1T)_N}\propto(M^*)^2\propto B^{-4/3}.
\end{equation}
Thus, we predict that the application of a magnetic field $B$
leads to a crossover from  NFL to LFL behavior and to a
significant reduction in both the relaxation rate and the
thermal resistivity, as the normalized LMR of $\rm YbRh_2Si_2$
does, see Fig. \ref{T12} B.

Our next step is analysis of the herbertsmithite low-frequency
optical conductivity $\overline{\sigma}$. To avoid the
contribution of phonon absorption into the conductivity, we
consider low temperatures $T$ and frequencies ${\omega}$
\cite{jltp:2018}. This is because the above contribution becomes
substantial at elevated $T$ and $\omega$ \cite{Pilon}. In the
case of QSL the optical conductivity is given by
\cite{jltp:2018}
\begin{equation}\label{sigma}
\overline{\sigma}(\omega)\propto\omega\chi{''}(\omega)\propto
\omega^2(M^*)^2.
\end{equation}
It follows from Eq.~\eqref{sigma} that
$\overline{\sigma}(\omega) \propto \omega^2$, and that behavior
is consistent with experimental facts obtained in measurements
on $\rm ZnCu_{3}(OH)_6Cl_2$ and $\rm EtMe_3Sb[Pd(dmit)_2]_2$
representing the best candidates for identification as a
material that hosts QSL \cite{Pilon,pust2018}. It is seen from
Eqs. \eqref{MTT} and \eqref{sigma}, that at elevated
temperatures the low-frequency optical conductivity is a
decreasing function of $T$. This observation is consistent with
the experimental data \cite{Pilon}, see also \cite{Lee}. It also
follows from Eqs. \eqref{MBB} and \eqref{sigma} that
$\overline{\sigma}(\omega)$ diminishes under the application of
magnetic fields. This observation seems to contradict the
experimental results since no systematic magnetic field
dependence is observed \cite{Pilon}. To elucidate the magnetic
field dependence of $\overline{\sigma}(B)$, we note that
measurements of $\overline{\sigma}(B)$ have been taken at 6 K
and the magnetic fields $B\leq 7$ T \cite{Pilon}. In such a case
the system is still in the transition regime and does not enter
into the LFL state at which the effective mass $M^*$ is given by
Eq. \eqref{MBB} \cite{prr}. Therefore, in this case the
effective mass behavior is determined by Eq. \eqref{MTT}, rather
than by Eq. \eqref{MBB}, and the $\overline{\sigma}(B)$
dependence cannot be observed. As a result, we predict that the
$B$-dependence of $\overline{\sigma}$ can be observed at
$B\simeq 7$ T provided that $T\leq 1$ K. In that case, as it
seen from Fig. \ref{fig02}, at $T\leq 1$ K the effective mass
$M^*\propto C_{\rm mag}/T$ is a diminishing function of the
applied magnetic field. Thus, we predict that
$\overline{\sigma}(B)$ diminishes at growing magnetic fields, as
it follows from Eqs. \eqref{MBB} and \eqref{sigma}. We note that
the above experiments on measurements of the heat transport and
optical conductivity can be carried out on samples with
different $x$. As a result, these experiments allow us to test
the influence of impurities on the value of the gap. We predict
that at moderate $x\sim 20$\% QSL remains robust, for both the
inhomogeneity and randomness facilitate frustration.

In summary, the main message of our paper is to suggest
performing the above discussed heat transport, low energy
inelastic neutron scattering, and optical conductivity
$\overline{\sigma}$ measurements on $\rm ZnCu_{3}(OH)_6Cl_2$
subjected to external magnetic fields. We have suggested that
growing $x$, characterizing \% of the $\rm Zn$ sites that are
occupied by $\rm Cu$, can facilitate the frustration of the
lattice, and thus can stabilize QSL. This observation can be
tested in experiments on samples of herbertsmithite with
different $x$. Considered measurements can give an unambiguous
answer if a real gap in spinon excitations, determining the
thermodynamic, transport and relaxation properties of insulating
magnets, exists, or does not exists, and how it depends on
impurities. Such measurements pave a new avenue for
technological applications of the magnets.

We are grateful to V.A. Khodel for valuable discussions. This
work was partly supported by U.S. DOE, Division of Chemical
Sciences, Office of Basic Energy Sciences, Office of Energy
Research.  JWC acknowledges support from the McDonnell Center
for the Space Sciences, and expresses gratitude to the
University of Madeira and its branch of Centro de
Investiga\c{c}\~{a}o em Matem\'{a}tica e Aplica\c{c}\~{o}es
(CIMA) for gracious hospitality during periods of extended
residence.

\end{document}